\documentstyle[aps,epsfig,multicol,fancyheadings]{revtex}
\begin{document}
%%%%%%%%%%%%%%%%%%%%%%%%%%%%%%%%%%%%%%%%%%%%%%%%%%%%%%%%

\title{Energy landscapes in random systems, driven interfaces and wetting}

\author{E.\ T.\ Sepp\"al\"a and M.\ J.\ Alava}

\address{Helsinki University of Technology, Laboratory of
Physics,\\ P.O.Box 1100, FIN-02015 HUT, Finland}

\maketitle

\begin{abstract}
We discuss the zero-temperature susceptibility of elastic manifolds
with quenched randomness. It diverges with system size due to
low-lying local minima.  The distribution of energy gaps is deduced to
be constant in the limit of vanishing gaps by comparing numerics with
a probabilistic argument.  The typical manifold response arises from a
level-crossing phenomenon and implies that wetting in random systems
begins with a discrete transition.  The associated ``jump field''
scales as $\langle h \rangle \sim L^{-5/3}$ and $L^{-2.2}$ for (1+1)
and (2+1) dimensional manifolds with random bond disorder.
\end{abstract}

\noindent {\it PACS \# \ \ 75.50.Lk, 05.70.Np, 68.45.Gd, 74.60.Ge}

\begin{multicols}{2}[]
\narrowtext

The physics of systems with quenched disorder is related to the energy
landscape. The free energy is at low temperatures governed by zero
temperature effects, which in turn are ruled by the scaling of the
disorder-dependent contribution.  Random magnets, as spin glasses and
random field systems, flux line lattices in superconductors, and
granular materials are examples of physical systems in which
frustration and disorder play an important role. Disorder may dominate
also in non-equilibrium conditions, like driven systems (domain walls
in magnets, flux lines in superconducting materials). In that case
temperature-driven dynamics (creep, aging) and the external drive
change the system from one metastable state to
another~\cite{blatter,review}.

A lot of information about energy landscapes is contained in how the
number of local energy minima and the typical scale of their energy
differences scale with system size, $L$ \cite{Hou1}. This can be
interpreted in a geometric fashion in that one compares the energy
difference of two states with their overlap in terms of the spin
configuration (as for magnets).  In spin glasses an intense debate
still goes on: whether in the thermodynamic limit the thermodynamic
state is trivial (``droplet'' picture~\cite{droplet}) or not (as in
the ``replica symmetry breaking'' picture~\cite{replica}).

Consider now the problem of the energetics of $D$ dimensional elastic
manifolds in random media \cite{Fis86,Emig99,HaH95,Las98}, of which
the best-known case is a directed polymer (DP) in a random medium with
$D=1$, often called a 'baby spin-glass'~\cite{Parisi90}.  For these
systems the interface energy is proportional to the area, and the
sample-to-sample energy fluctuations scale with the exponent $\theta$,
($\theta =$ 1/3 for a DP in $d=D+1=2 $ embedding dimensions). The
geometry is often self-affine, characterized by a roughness exponent
$\zeta$, (2/3, when $d=2$).  In the simplest energy landscape the
valleys and excitations are separated by energy gaps proportional to
$l^\theta$ where $l$ is the length scale of the perturbation.

Here the susceptibility of elastic manifolds is studied in the
presence of weak fields numerically and by scaling arguments.  By
investigating each sample separately, we explore the changes in the
energy landscape with applied fields.  These lead to discrete 'jumps'
in the physical configuration.  As a consequence scaling arguments of
wetting in random systems do not work in the limit of weak fields if
the original interface-to-wall distance is much larger than the
interface roughness~\cite{Lip86}.  With pre-conditioned systems we
obtain the detailed probability distribution of the energy differences
(gaps) between local minima and the global one. We find that the
average interface behavior can be explained with scaling arguments,
but the susceptibility can not, and it is directly related to the
exact properties of the gap distribution.  Thus the detailed
statistics of the landscape is important. This contradicts
considerations for random systems that assume well-defined
thermodynamic functions~\cite{Shapir91} and scaling arguments with a
single parameter ($L^\theta$).  These findings agree with claims that
the susceptibility of a DP to thermal perturbations or applied fields,
is anomalous \cite{Mez90,Hwa94,Fish91}.  The reason is that the
response to a very weak field, say applied locally at the end-point of a DP, is
governed by rare samples.  The disorder-averaged response differs from
the typical one because the ground state can be almost degenerate with
a local minimum.  Likewise, numerical studies of $d = (1+1)$ DP
susceptibility reveal aging phenomena reminiscent of real
spin-glasses~\cite{Yoshino,Barrat97}.

The continuum Hamiltonian for a $D$ dimensional elastic manifold 
(${\bf x}$  is an internal coordinate and a $z$ (scalar)
displacement)
\begin{equation}
{\mathcal H} = \int {\rm d}^D{\bf x} \, \left[ \Gamma \{ \nabla 
z({\bf x}) \}^2 + V_r ({\bf x},z) + h(z) \right],
\label{H}
\end{equation} 
with an elastic energy ($\Gamma$ is the interface stiffness), and
$V_r$ a random pinning energy (we use a random bond correlator,
$\langle V_r ({\bf x},z) V_r ({\bf x'},z') \rangle = 2 {\mathcal D}
\delta({\bf x}- {\bf x}') \delta(z-z')$).  $h(z)$ couples the
interface to an external perturbation, e.g.\ it describes a constant
magnetic field $H$ in Ising magnets with antiperiodic boundary
conditions.

The Hamiltonian (\ref{H}) describes also complete wetting in a random
system, where $h(z)$ equals to the chemical potential difference of
the wetting layer and the bulk phase
\cite{Lip86,Hua89,Wuttke91,Dole91}.  For $h$ non-negligible the
wetting-inducing external potential competes with the tendency of the
interface to win pinning energy. Assuming that these balance, the
average interface-wall separation $\langle z \rangle$ becomes $\langle
z \rangle \sim h^{-\psi}, \psi = \frac{1}{\tau + \kappa}$ where $\psi$
is the depinning exponent. $\tau$ measures the scaling of the elastic
and pinning energy and is given by $\tau = 2 (1-\zeta) / \zeta$, and
$\kappa$ is the scaling exponent of the external field $h(z) \sim
z^\kappa$ (here we use $\kappa =1$).  For random bond systems $\tau=1$
in $d=1+1$ dimensions, and $\tau\simeq$ 2.9 in $d=2+1$ using the known
bulk roughness exponent values $2/3$ and 0.41 in $d=2$ and 3,
respectively \cite{Fis86,Alava96}. In $d=2$ numerical simulations in
random Ising systems indicate, in agreement, $\psi \simeq 0.5$
\cite{Hua89,Wuttke91}.

\begin{figure}[f]
\centerline{\epsfig{file=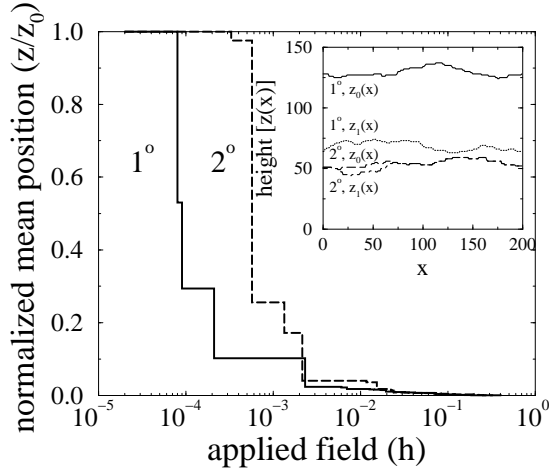,width=7cm,angle=-90}}
\caption{Overview of two realizations of changes in mean heights
$\bar{z}$ of interfaces normalized by their original (global minimum)
positions $\bar{z}_0$ vs. applied field $h$ for $(1+1)$ dimensional
systems. Note the large jumps in both cases.  $L^2=200^2$. $J_{ij,z}
\in [0,1]$ uniform distribution and $J_{ij,x} = 0.5$ (random bond
disorder).  The expected scenarios (bubble formation, jump to the
lower edge of the system) before and after the first moves from global
minima $z_0(x)$ to $z_1(x)$ are shown in the inset.}
\label{fig1}
\end{figure}

A network flow algorithm, invented by Goldberg and
Tarjan~\cite{Goltar88}, is used here for the numerical procedure.  It
solves the minimum cut - maximum network flow problem, and produces in
polynomial time the exact ground state energy and interface
configuration given a sample ($L \times L_z$ or $L \times L \times
L_z$) with fixed quenched disorder. $L_z$ is the $z$-directional
system size. The algorithm is convenient when one makes systematic
perturbations to the original problem $(h=0)$ \cite{Alavaetal,Joe98}.
Figure \ref{fig1} illustrates the sample-to-sample behavior, as the
external field $h(z)$ is switched on slowly (see Eq.~(\ref{H})).  At
$h=0$ the interface is in the ground state. It has a mean wall
distance $\bar{z}_0$ and a width $w \sim L^\zeta$ in a system of
transverse size $L_z$. As the field is increased the interfaces move
intermittently with jumps to positions ($\bar{z}_1, \bar{z}_2, \ldots,
\bar{z}_n, \ldots$) \cite{comm1}. This corresponds to a {\it
first-order transition}.  Instead of finite-size excitations the first
change in the interface configuration is a macroscopic jump with zero
overlap between the old and new states. The first transition point
defines a jump field $h_1$. It assumes the role of a latent heat, and
corresponds to the landscape-dependent energy to move the interface.

The two possible mechanisms are compared in the inset of
Fig.~\ref{fig1}.  Either the interface adjusts itself gradually by
forming 'bubbles' or local excitations, or it jumps completely
(compare with the main figure). The scenarios are linked to the
structure of the energy landscape. If the first excitation is
localized and has the transverse spatial extension $\Delta$ ($l \simeq
\Delta^{1/\zeta}$)~\cite{Hwa94}, the energy cost scales with
$\Delta^{a/\zeta}$ and the energy win in the field scales with $h_1
\Delta^{1+(d-1)/\zeta}$.  Assuming that $a =\theta$ the jump field
$h_1 \sim \Delta^{\bar{\alpha}} =\Delta^{\theta/\zeta -1 - (d-1)
/\zeta}$. The exponent is negative, and thus small excitations are the
more expensive ones \cite{finiteT}. Numerically, the fraction of jumps 
leading to a
non-zero overlap with the ground state decreases towards zero slowly 
with $L$. Also, the scaling function of the interface jump lengths
approaches a constant shape. The mean jump length ($\Delta z_1 =
\bar{z}_0 - \bar{z}_1$, $\bar{z}_1 < \bar{z}_0$) scales extensively,
$\Delta z_1 \sim L_z$, not with e.g. $L^\zeta$.

\begin{figure}[f]
\centerline{\epsfig{file=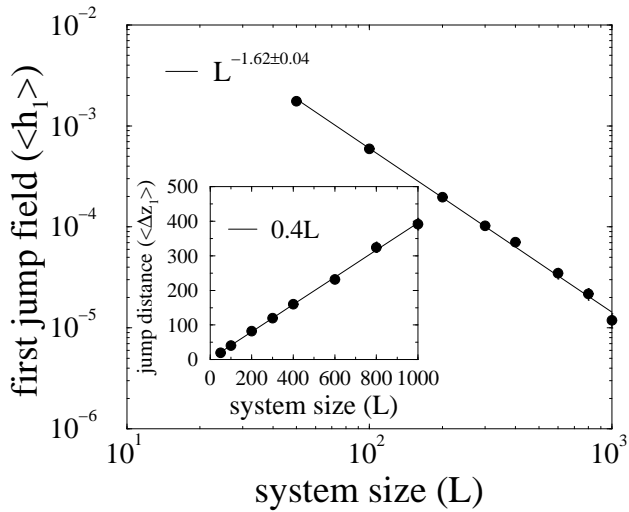,width=7cm,angle=-90}}
\caption{Finite size scaling of the average first jump field $\langle
h_1 \rangle$ for one dimensional DP's. The line is the least squares
fit to data. The scaling argument gives $\alpha = 5/3$.  The inset
shows the average jump distance $\langle \Delta z_1 \rangle$ at the
corresponding field $h_1$ with a linear fit to data.  $\langle \,
\rangle$ is the disorder-average over $N=1000$ realizations for the
system sizes $L \times L_z =L^2 = 50^2$ and $100^2$, $N=500$ for $L^2
= 200^2\--400^2$, and $N=200$ for $L^2 = 600^2\--1000^2$.  The
disorder is of random bond type.}
\label{fig2}
\end{figure}

So for small fields $h$ and $L^\zeta \ll L_z$ the
sample-to-sample fluctuations lead to a {\it discrete} (wetting)
transition. The average behavior with $\langle z(h) \rangle$ and
typical interface behavior with $\bar{z}(h)$ do not coincide, since
the asymptotic $h \rightarrow 0$ limit is dominated by the
near-degeneracy of the ground state.  In the limit $L^\zeta \ll L_z$
there are many independent 'valleys' in the energy landscape for
directed surfaces. Each of these has an energy $E_n$ corresponding to
a local minimum and their energy difference to the ground state (with
$E_0$) is expected to scale as with two independent sets of
disorder. That is $E_n - E_0 \sim L^\theta$.  This energy difference
equated with the jump energy $h_1 L^D \Delta z_1$ leads (with the
choice $L_z =L$) to the scaling
\begin{equation}
h_1 \sim L^{\theta-d} = L^{-\alpha}.
\end{equation}
The jump field exponents are $\alpha = 5/3$ and $\alpha \simeq 2.18$
in $d=2$ and $d=3$ random bond systems, respectively. In $d=3$ random
field interfaces have $\alpha = 5/3$ ($\zeta = 2/3$ and $\theta = 2
\zeta +D - 2$ \cite{Emig99,HaH95}).  It is assumed that $\Delta z_1
\sim L$, since the valley energies are independent, except for the
bias caused by the field $h$.  Figure \ref{fig2} compares the exponent
values to numerical data with only the non-overlapping jumps being
considered (without this pruning the same exponent is obtained
asymptotically). For $D=1$ $\alpha$ becomes $1.62 \pm 0.04$, close to
the scaling estimate of 5/3. The inset shows the disorder-averaged
jump distance $\langle \Delta z_1 \rangle$ vs. $L$ and shows that the
interface response geometry scales linearly with $L$ (as discussed
above). For $D=2$ random bond manifolds we obtain $\alpha \simeq 2.2$,
in reasonable agreement again.  In the limit $\langle z_n(h) \rangle
\simeq \bar{z}_n(h) \simeq w \sim L^\zeta$ (after $n$ jumps of sizes
$\Delta z_n = \bar{z}_{n-1} - \bar{z}_n$) the mean-field wetting
theory applies, and indeed we obtain for the depinning exponent for
$d=2$ $\psi \simeq 1/2$, and for $d=(2+1)$ $\psi \simeq 0.26$, in
rough accordance with the Lipowsky-Fisher~\cite{Lip86} prediction.  In
$d=(2+1)$ there are deviations including a dewetting transition for
weak disorder \cite{Alava96} and the exponent converges very slowly
($\bar{z}_0 \simeq w \sim L^\zeta$ at $L \simeq 10^4$ if $L_z = 50$
\cite{pitkapaperi}).

\begin{figure}[f]
\centerline{\epsfig{file=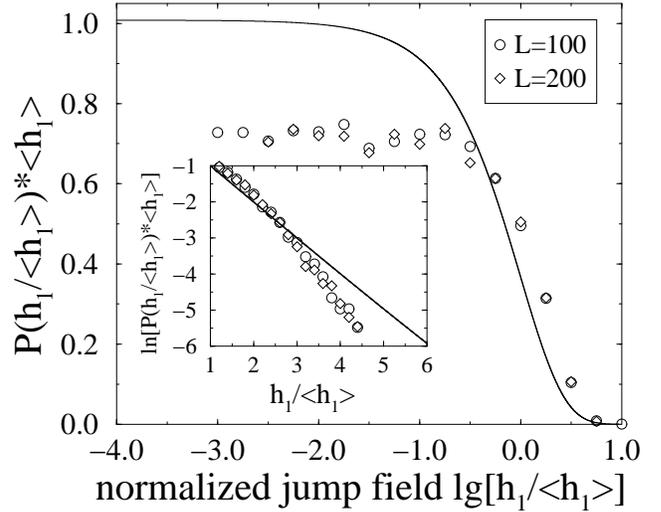,width=7cm,angle=-90}}
\caption{The scaling function of the probability distribution $P(h_1/
\langle h_1 \rangle)\times \langle h_1 \rangle$ for the first jump
field values $h_1$ normalized by their disorder-average $\langle h_1
\rangle$ in a (10-base) semilog-scale for the system sizes $L \times
L_z = L^2 =100^2$ and $200^2$. The inset shows the tails in the
natural-log-scale. The initial global minimum position $\bar{z}_0/L_z
= \rm{const}$ for all $L$. The number of realizations $N=10^4$ for
both system sizes. The line is the analytic result from
Eq.~(\ref{analytic}) with a uniform distribution $\hat{P}(x)$ and
$N_z=20$.}
\label{fig3}
\end{figure}

\begin{figure}[f]
\centerline{\epsfig{file=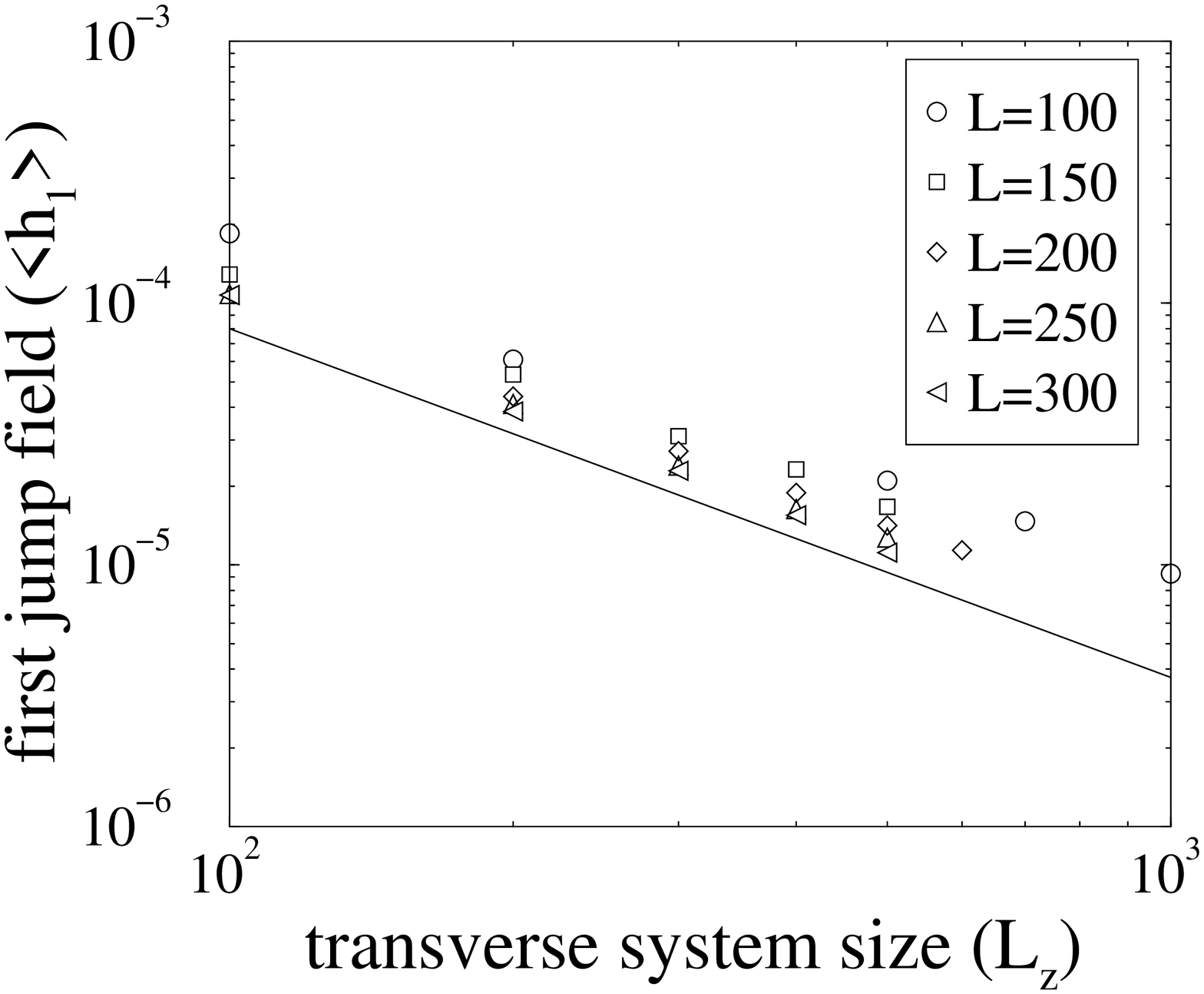,width=7cm,angle=-90}}
\caption{The disorder-average of the first jump field $\langle h_1
\rangle$ as a function of transverse system size $L_z$ for the system
sizes $L=$100, 150, 200, 250 and $300$, each with $\bar{z}_0/L_z =
\rm{const}$. The number of realizations ranges from $N=500$ for
$L=300$, $L_z=500$ to $N=2600$ for $L=200$, $L_z=600$. The line
$L_z^{-\gamma}, \gamma =4/3$ is a guide to the eye.}
\label{fig4}
\end{figure}

If the initial interface position is random, the jump statistics are
an average over the initial number of available valleys (recall that
the field breaks the up-and-down-symmetry, see Fig.~\ref{fig1}).  Thus
we also consider the limit in which the initial position is set to be
inside a fixed-size window, $\bar{z}_0/L_z \simeq \mathrm{const}$.  We
expect that the number of local valleys in the landscape, accessible
with $h>0$, has a well-defined average (in the grand-canonical sense),
and that the relevant scaling parameter is $L_z / L^\zeta$. Figure
\ref{fig3} shows the scaling function of the probability distribution
$P(h_1)$ obtained with this initial condition.  We find the form
$P(h_1/\langle h_1 \rangle) = A(L) f(h_1/\langle h_1\rangle)$ where
$A$ depends on the energy gap scale $L^\theta$ and $f$ is a scaling
function with the limiting behaviors $f(x) \rightarrow 1, x
\rightarrow 0$ and $f(x) \sim \exp{(-ax^\beta)}, x > 1, \beta \simeq
1.3$.  The distribution is constant for small fields and has an almost
exponential cut-off.  The scaling properties imply in particular that
the disorder-averaged susceptibility {\it diverges}.  The change in
magnetization is given by the number of interfaces that have moved
times the mean distance $\langle \Delta z_1 \rangle$.  Thus the divergence
is not $\chi_{tot} \sim L^3$ \cite{Shapir91}.  Figure \ref{fig4}
shows the average jump field in the fixed height ensemble with varying
$L_z$ and constant $L$. We have fitted the data with $\langle h_1
\rangle \sim L_z^{-\gamma}$, and the best fit is obtained by the
scaling exponent $\gamma \simeq 4/3$.

Consider now the energy landscape for small $h$. It has $k = 1, \dots,
N_z$ associated minima ($N_z \sim L_z / L^\zeta$) with the energies
$E_k$ picked out of an associated energy gap probability distribution
$\hat{P}(\Delta E_k)$, where $\Delta E_k = E_k - E_0$ and $E_0$ is the
ground state energy. When $h > 0$, all the local minima attain an
energy of $E_k + h \Delta z_k$ with respect to the reference state
with $\bar{z}_0$ and $E_0$. Now we make the assumption, analogous to
the Random Energy Model \cite{derrida}, that all the gap energies
$\Delta E_k$ are independent random variables. We can now simply
compute the probability for the original ground state being stable for
any $h$ (i.e. no jump has taken place) by the joint probability $P_0$
that all the $E_k + h\Delta z_k$'s are still higher than the original
one with the given $h$. $-\partial P_0/ \partial h$ gives then the
probability that this {\it level crossing} occurs at exactly $h$. By
computing
\begin{equation}
\frac{\partial P_0}{\partial h} = - {\rm e}^{-\int_1^{N_z} \int_0^{kh/N_z}
\hat{P}(x) \, dx \, dk} \int_1^{N_z} \frac{ \hat{P}(kh/N_z)}{1 -
\int_0^{kh/N_z} \hat{P}(x) \, dx} \, dk
\label{analytic}
\end{equation} 
one can show that the only $\hat{P}$ that reproduces the numerical
$P(h_1)$ is a {\it constant} one, whereas all other functional forms
of $\hat{P}$ fail, see Fig.~\ref{fig3}. This $\hat{P}$ is in fact
exactly the marginal one needed for the susceptibility per spin $\chi
= \lim_{h \rightarrow 0} \langle \partial \bar{z}/ \partial h \rangle$
to diverge in the thermodynamic limit. In particular for a
distribution $P(h_1)$ that vanishes in the zero field limit the
susceptibility would stay finite.  Using the obtained form for the
probability distribution gives $\chi \sim L^\theta
\left(\frac{L_z}{L^\zeta} \right)^\gamma$ where $\gamma \simeq
1-\zeta$ relates to the density of valleys. This slightly disagrees
with the above result ($\gamma \simeq 4/3$) since with $L =
\mathrm{const}$ $\chi \sim L_z^\gamma$, $\gamma = 1$.  In the
isotropic limit $L \propto L_z$ the extensive susceptibility simply
reads $\chi_{tot} = L^d \chi \sim L^{d+1+\theta-\zeta} \sim
L^{2D+\zeta}$. To conclude, $\chi$ (or $\chi_{tot}$) is determined by
the exact low-energy properties of $\hat{P}$, or by the rare events in
the low $\Delta E$ tail.

To summarize we have studied the coupling between the energy landscape
structure and the response of interfaces, related for instance to
complete wetting. A disorder averaging that reflects correctly the
level-crossing character of the problem reveals that the wetting
starts with a discrete transition. Thus the randomness of the energy
landscape drives a second-order transition to a first-order one. The
'jump' is associated with an effective specific heat, which can be
understood in terms of scaling arguments. The susceptibility is
governed by the infrequent cases with low-lying local minima, which
allows us to derive a constant energy gap probability distribution.
The results should be relevant for other problems like flux line
lattices in superconducting materials with quenched
randomness~\cite{blatter}. It will also be of interest to see if the
energetics and the geometrical character of the response can be
coupled with arguments concerning the energy barriers in each specific
configuration \cite{vinokur}. This would allow to understand the
dynamics in the creep regime, when the interface moves between
metastable states.

Phil Duxbury is acknowledged for a crucial suggestion, and Simone Artz,
Martin Dub\'e, and Heiko Rieger for discussions.  We thank 
the Academy of Finland for support.

%%%%%%%%%%%%%%%%%%%%%%%%%%%%%

%%%%%%%%%%%%%
%FIGURES
%%%%%%%%%%%%%

\end{multicols}


\begin{references}
\bibitem{comm1} Configurational changes are determined with the numerical
accuracy $\Delta h = 10^{-5}$.
\bibitem{finiteT} The argument about preferred large-scale excitations
can not be used for $T>0$, though the level-crossing between
valleys should be true for the free-energy at finite temperatures.
\end{references}
\end{document}